\documentclass[%
reprint,
superscriptaddress,
amsmath,
amssymb,
aps,
floatfix,
longbibliography
]{revtex4-1}

\usepackage{amsmath}
\usepackage{graphicx}
\usepackage{units}
\usepackage{braket}
\usepackage{float}
\usepackage{csquotes}
\usepackage{array}
\usepackage[colorlinks, linkcolor=blue, citecolor=blue, urlcolor=blue, breaklinks=true]{hyperref}

\usepackage[dvipsnames]{xcolor}
\usepackage[normalem]{ulem}

\newcommand{\Tr}{\operatorname{Tr}}
\newcommand{\ketbra}[2]{\left\vert#1 \right\rangle\left\langle#2\right\vert}

\begin{document}

\preprint{APS/123-QED}

\title{Non-local temporal interferometry for highly resilient free-space quantum communication}

\author{Lukas Bulla}
\email{lukas.bulla@oeaw.ac.at}
\affiliation{Institute for Quantum Optics and Quantum Information (IQOQI), Austrian Academy of Sciences, Boltzmanngasse 3, 1090 Vienna, Austria}
\affiliation{Vienna Center for Quantum Science and Technology (VCQ), Faculty of Physics, University of Vienna, Boltzmanngasse 5, 1090 Vienna, Austria}

\author{Matej Pivoluska}
\email{mpivoluska@mail.muni.cz}
\affiliation{Institute for Quantum Optics and Quantum Information (IQOQI), Austrian Academy of Sciences, Boltzmanngasse 3, 1090 Vienna, Austria}
\affiliation{Vienna Center for Quantum Science and Technology (VCQ), Faculty of Physics, University of Vienna, Boltzmanngasse 5, 1090 Vienna, Austria}
\affiliation{Institute of Computer Science, Masaryk University, 602 00 Brno, Czech Republic}
\affiliation{Institute of Physics, Slovak Academy of Sciences, 845 11 Bratislava, Slovakia}

\author{Kristian Hjorth}
\affiliation{Institute for Quantum Optics and Quantum Information (IQOQI), Austrian Academy of Sciences, Boltzmanngasse 3, 1090 Vienna, Austria}
\affiliation{ 
Department of Physics, Faculty of Natural Sciences, Norwegian University of Science and Technology (NTNU), NO-7491 Trondheim, Norway}

\author{Oskar Kohout}
\affiliation{Institute for Quantum Optics and Quantum Information (IQOQI), Austrian Academy of Sciences, Boltzmanngasse 3, 1090 Vienna, Austria}
\affiliation{Fraunhofer Institute for Applied Optics and Precision Engineering IOF,Albert-Einstein-Strasse 7, 07745 Jena, Germany}
\affiliation{Friedrich-Schiller-Universität Jena FSU, Fürstengraben 1, 07743 Jena, Germany}

\author{Jan Lang}
\affiliation{Institute for Quantum Optics and Quantum Information (IQOQI), Austrian Academy of Sciences, Boltzmanngasse 3, 1090 Vienna, Austria}
%\affiliation{Vienna Center for Quantum Science and Technology (VCQ), Faculty of Physics, University of Vienna, Boltzmanngasse 5, 1090 Vienna, Austria}

\author{Sebastian Ecker}
\affiliation{Institute for Quantum Optics and Quantum Information (IQOQI), Austrian Academy of Sciences, Boltzmanngasse 3, 1090 Vienna, Austria}
\affiliation{Vienna Center for Quantum Science and Technology (VCQ), Faculty of Physics, University of Vienna, Boltzmanngasse 5, 1090 Vienna, Austria}

\author{Sebastian P. Neumann}
\affiliation{Institute for Quantum Optics and Quantum Information (IQOQI), Austrian Academy of Sciences, Boltzmanngasse 3, 1090 Vienna, Austria}
\affiliation{Vienna Center for Quantum Science and Technology (VCQ), Faculty of Physics, University of Vienna, Boltzmanngasse 5, 1090 Vienna, Austria}

\author{Julius Bittermann}
\affiliation{Institute for Quantum Optics and Quantum Information (IQOQI), Austrian Academy of Sciences, Boltzmanngasse 3, 1090 Vienna, Austria}
\affiliation{Vienna Center for Quantum Science and Technology (VCQ), Faculty of Physics, University of Vienna, Boltzmanngasse 5, 1090 Vienna, Austria}

\author{Robert Kindler}
\affiliation{Institute for Quantum Optics and Quantum Information (IQOQI), Austrian Academy of Sciences, Boltzmanngasse 3, 1090 Vienna, Austria}
\affiliation{Vienna Center for Quantum Science and Technology (VCQ), Faculty of Physics, University of Vienna, Boltzmanngasse 5, 1090 Vienna, Austria}

\author{Marcus Huber}
\email{marcus.huber@tuwien.ac.at}
\affiliation{Institute for Quantum Optics and Quantum Information (IQOQI), Austrian Academy of Sciences, Boltzmanngasse 3, 1090 Vienna, Austria}
\affiliation{Vienna Center for Quantum Science and Technology (VCQ), Atominstitut, Technische  Universit{\"a}t  Wien,  Stadionallee 2, 1020  Vienna,  Austria}

\author{Martin Bohmann}
\email{martin.bohmann@oeaw.ac.at}
\affiliation{Institute for Quantum Optics and Quantum Information (IQOQI), Austrian Academy of Sciences, Boltzmanngasse 3, 1090 Vienna, Austria}
\affiliation{Vienna Center for Quantum Science and Technology (VCQ), Faculty of Physics, University of Vienna, Boltzmanngasse 5, 1090 Vienna, Austria}

\author{Rupert Ursin}
\email{rupert.ursin@oeaw.ac.at}
\affiliation{Institute for Quantum Optics and Quantum Information (IQOQI), Austrian Academy of Sciences, Boltzmanngasse 3, 1090 Vienna, Austria}
\affiliation{Vienna Center for Quantum Science and Technology (VCQ), Faculty of Physics, University of Vienna, Boltzmanngasse 5, 1090 Vienna, Austria}

\date{\today}

\begin{abstract}
Entanglement distribution via photons over long distances enables many applications, including quantum key distribution (QKD), which provides unprecedented privacy. The inevitable degradation of entanglement through noise accumulated over long distances remains one of the key challenges in this area.
Exploiting the potential of higher-dimensional entangled photons promises to address this challenge, but poses extreme demands on the experimental implementation.
 Here, we present an interstate free-space quantum link, distributing hyper-entanglement over $10.2\,$km with flexible dimensionality of encoding by deploying a phase-stable non-local Franson interferometer.
With this distribution of multidimensional energy-time entangled photons, we analyse the achievable key rate in a dimensionally-adaptive QKD protocol that can be optimized with respect to any environmental noise conditions.
Our approach enables and emphasises the power of high-dimensional entanglement for quantum communication, yielding a positive asymptotic key rate well into the dawn of the day.
\end{abstract}

\maketitle

\begin{figure*}[ht]
        \centering
        \includegraphics[width=2\columnwidth]{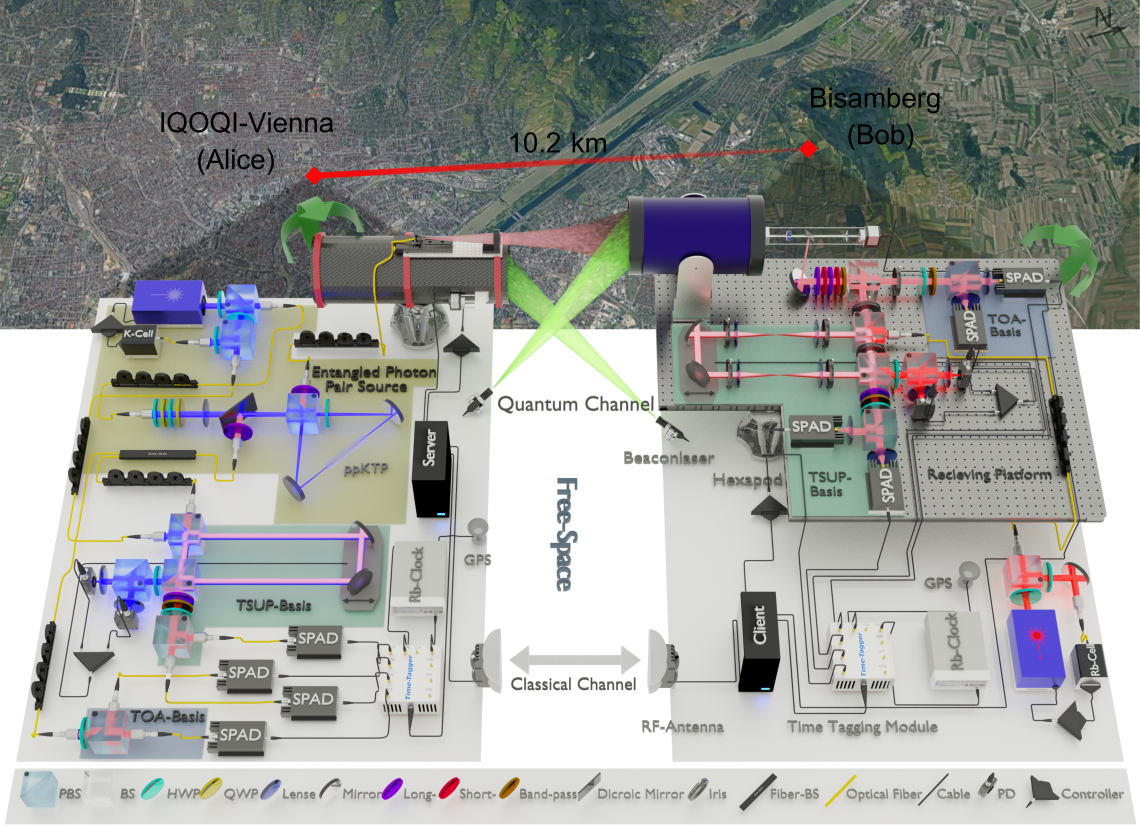}
        \caption{
        Illustration of the high-dimensional quantum key distribution experiment.
        The free-space link was established between the sender at IQOQI Vienna (Alice) and the receiver at Bisamberg (Bob) in Lower Austria over a distance of $10.2$~km.
        The hyperentangled photon source is pumped by a $^{39}K$ stabilized Laser at $404.532$ nm; 
        The hyperentangled state was set by adjusting a combination of Half-Wave (HWP) and Quater-Wave plates (QWP); The hyperentangled photons produced at $808.9$ nm by a ppKTP crystal are separated and guided to the receivers Alice and Bob who by the aid of a 50:50 beam splitter (BS) randomly measure their photons in the time-of-arrival (TOA) and the temporal superposition (TSUP) basis; By using polarizing beam splitters (PBS) in the Mach-Zehnder-Interferometers (MZI) followed by a polarization measurement in the D/A-basis we achieve post-selection free Franson interference. While Alice analyzes her photons locally, Bob's photons are transmitted over the free-space link to Bisamberg. The MZI is locked to the pump laser stabilized by a piezo actuator controlled by the interferometric intensity changes measured by two photo-detectors (PD).
        The photon detection events are recorded by single-photon avalanche diodes (SPAD), time stamped by the time-tagging module and streamed to the server.
        The photons received at Bob are guided to his detection modules where we used several irises as optical baffles and combinations of optical filters to reduce the influence of background radiation.
        Bob's measurements are corresponding to Alice's, with the difference of having two 4f-systems in the MZI to compensate for the atmospheric turbulence; Bobs MZI is locked to a $^{87}Rb$ stabilized laser at $780.23$ nm  }
        \label{fig:linkv1}
    \end{figure*}
\section{\label{sec:Introduction}Introduction}

    Securing communication against eavesdropping is one of the most important challenges of modern society with its constantly increasing use in information technology.
    The security of our current encryption methods, however, is challenged by the computational power of emerging quantum computers and vulnerable to device hacking.
    A solution to these threats can also be found in the quantum realm, namely by means of quantum key distribution (QKD) \cite{gisin2002,xu2020}.
    In contrast to classical cryptography relying on computational hardness assumptions, the security of QKD is based on the very nature of physical laws.
    In particular, quantum entanglement \cite{horodecki2009} allows to establish a secure cryptographic key between two communicating parties, which can be used to establish a secure communication channel. 
    Security of a cryptographic key produced by an entanglement-based QKD protocol is assured by the monogamy of quantum entanglement.
    This concept can be extended to multi-user communication networks \cite{wengerowsky2018entanglement,jin2018demonstration,marcikic200450kmtimbin,honjo2008} and thus has a great potential to revolutionize the way secrecy of communication is protected.
    Over the last decade, laboratory proof-of-principle experiments were turned into real-world entanglement-based QKD implementations over deployed fiber \cite{korzh2015provably,yin2016measurement}, horizontal free-space \cite{ursin2007,fedrizzi2009,wang2013,nauerth2013,steinlechner2017distribution}, and satellite-earth links \cite{liao2017satellite,yin2017satellite,takenaka2017,liao2017,yin2020}.
    Nevertheless, there are fundamental challenges to overcome in free-space QKD communication channels.
    Among these are long communication distances that cause high signal loss. With increasing loss a point appears where intrinsic dark counts and the presence of high intrinsic background count rates, which generally make QKD during the day considerably more difficult, become dominant.
    To overcome these limitations it is possible to exploit fundamental quantum correlations even further by leveraging properties of high-dimensional quantum states.
    This can be achieved through hyper-entanglement using simultaneous entanglement in different degrees of freedom (DOF) \cite{kwiat1997,barreiro2005} and using DOF that intrinsically offer high-dimensional entanglement such as the time-energy DOF \cite{PhysRevA.96.040303,martin2017quantifying,cuevas2013long}.
    Indeed, such high-dimensional quantum states do not only offer an increased quantum communication capacity \cite{bechmann-pasquinucci2000,cozzolino2019,zhong2015photon,Ali-KhanPhysRevLettLarge-Alphabet}, but also provide improved noise robustness in entanglement distribution \cite{ecker2019overcoming,nape2021measuring}.
    These remarkable features can be exploited in a QKD protocol with high-dimensional encoding \cite{doda2021quantum}. Although these features have been analysed in theory using simplified noise models and later also demonstrated under laboratory conditions \cite{hu2020}, a fully fledged demonstration of the benefits of high-dimensional quantum states over a real-world quantum channel is missing until today.
    
    In this paper, we implement the first high-dimensional entanglement-based real-world free-space link  and demonstrate its practical usefulness by computing the asymptotic key rate in the subspace encoding QKD  protocol \cite{doda2021quantum}.
    A significant innovation of this approach is that it enables an adaptive and flexible encoding, allowing for an optimized QKD performance even under very harsh environmental conditions.     By establishing a $10.2\,$km-long free-space quantum communication channel over Vienna's metropolitan area we expose the quantum channel to the major challenges of free-space quantum communication and show that our schemes help under all circumstances where noise plays, or start to play, a dominant role.
    To this end, we develop and advance techniques that will enable fundamental and applied experiments based on phase-stable non-local interference over free space and long distances. Our approach provides a pathway that harnesses the advantages of high-dimensional quantum correlations to overcome the major challenges of real-world quantum communication, paving the way for a resilient and resource-efficient quantum internet.

\section{\label{sec:Results}Results}

\subsection{Experiment/Implementation}

    To exploit the high-dimensional nature of the time-energy DOF, we produced hyperentangled photon pairs and measured two-photon interference in a Franson-type non-local interferometer \cite{franson1989bell} over an interstate free-space link.
    The interstate experiment was carried out between the sending party at IQOQI Vienna (Alice) and the receiving party (Bob), located at a $10.2$~km free-space distance at Bisamberg in Lower Austria (see Fig. \ref{fig:linkv1}).

    The free-space quantum channel runs north-northeast (bearing $10.89^{\circ}$), mainly across Vienna's metropolitan area and over the Danube river to Bisamberg. The strong influence of atmospheric turbulence on the signal propagating over such an urban environment led to fluctuation in the angle of arrival of $150$\,\textmu rad  and an average channel loss of $\sim 25$ dB, comparable to low Earth orbit (LEO) satellite links \cite{gruneisen2021}.
    
    We used spontaneous parametric down-conversion (SPDC) to create hyperentangled photon pairs at Alice's laboratory.
    The entangled-photon-pair source was operated in a Sagnac configuration \cite{kim2006phase,fedrizzi2007wavelength} bidirectionally pumped by a $404.53\,$nm continuous-wave laser producing photon pairs at $809\,$nm, with a coherence time of  $\tau_{\mathrm{p}} \sim 3\,$ps defined by the ppKTP crystal length (see Appendix \ref{sub:Methodsource}).
    The resulting quantum state features both time-energy and polarization entanglement and can be approximated by
    \begin{align}\label{eq:state}
          \ket{\Psi}_{\mathrm{AB}} =\int dt\, f(t) \ket{t,t}
          \otimes
          \left( \ket{H,H} +e^{-i\phi} \ket{V,V}\right)
    \end{align}
    where $f(t)$ is a continuous function of time, determined by the coherence time of the laser, and H(V) indicates the horizontal (vertical) polarization state.
    One photon of a pair was detected locally at Alice's laboratory while the other was transmitted over the turbulent free-space link to Bob. The Transmitter was composed of a $75\,$mm achromatic sending lens and the beacon receiving $350\,$mm Newtonian telescope  mounted on top of a hexapod. Bob's receiver consisted of a Cassegrain telescope with a  $254\,$mm aperture also mounted on a hexapod in addition to the entire receiving module.
    To achieve consistent long-time-averaged count rates over the free-space link we implemented a bidirectional angular tracking of $532\,$nm beacon lasers at the transmitter and the receiver \cite{ursin2007}.

    Each communicating party was equipped with a receiving module capable of performing measurements in two  incompatible bases: the time-of-arrival basis (TOA) and the time-superposition (TSUP) basis (see Fig. \ref{fig:linkv1}). To measure in TSUP basis, the first-ever reported non-local Franson-type interferometer \cite{franson1989bell} separated by a distance of $10.2\,$km free-space was build.
    In the TOA basis, we detected the arrival time of the photons as well as their polarization (HV measurement), effectively decreasing the accidental noise photons as compared to a polarization insensitive bucket detector. 

    To perform the TSUP-basis measurements, we implemented unbalanced Mach-Zehnder interferometers (MZI) with a path difference of $\tau_{\mathrm{MZI}} = 2.7\,$ns, which exceeds the coherence length of the SPDC photons $(\tau_\mathrm{p} \ll  \tau_{\mathrm{MZI}} )$.
    
    By using polarizing beam splitters in the MZIs, we measured postselection-free Franson interference by mapping the polarization entanglement to the interferometer paths followed by a polarization measurement in the D/A-basis, effectively erasing the which-path information \cite{steinlechner2017distribution}.
    Thus, we exploited the hyperentanglement of the photon pairs to improve the time-energy TSUP measurement by means of polarization entanglement (see Appendix \ref{sub:MethodDetectionModules}).
    
    To enable non-local interference over a long-distance free-space link, which has to be phase-stable over significant times, we devised and implemented several experimental features.
    To counteract the detrimental effect of atmospheric turbulence and reduce mode mismatch between long and short interferometer path, caused by the difference in angles of arrival, we placed two 4f-systems in the long path of  Bob's MZI \cite{jin2018demonstration,jin2019genuine}.

    To reach the required high sub-wavelength demands in terms of phase stability and to avoid phase drifts due to thermal expansion of the MZIs, we stabilized them at the sender and the receiver by using two lasers locked to atomic hyperfine structure transitions \cite{debs2008laserlocking}, achieving frequency fluctuations below $1\,$MHz.
    By using the pump laser as stabilization laser we additionally locked the phase $\phi$ of the state \eqref{eq:state} to the phase of both interferometers (see Fig. \ref{fig:linkv1}).

    An essential point for using the time-energy DOF over long distances is to keep the reference time-frames of Alice and Bob synchronized, which is demanding due to relative drifts of the clocks and the small temporal size of time-bins.
    In order to solve this problem, we used GPS-synchronized $Rb$-clocks on both sides to lock the time-tagging modules for long-term stability.
    The time-tagging modules provided timestamps for each of the eight photon-detection events and streamed the data to the client and server data storage, respectively. 
    Furthermore, it was necessary to  compensate for relative short-term clock drifts $T_{\mathrm{dev}}\sim 30\,$ps/s by tracking via the statistical  photon intensity correlation function to gain accuracy \cite{Ho_2009,spiess2021clock}.

\begin{figure}[ht]
   \centering
  \includegraphics[width=1\columnwidth]{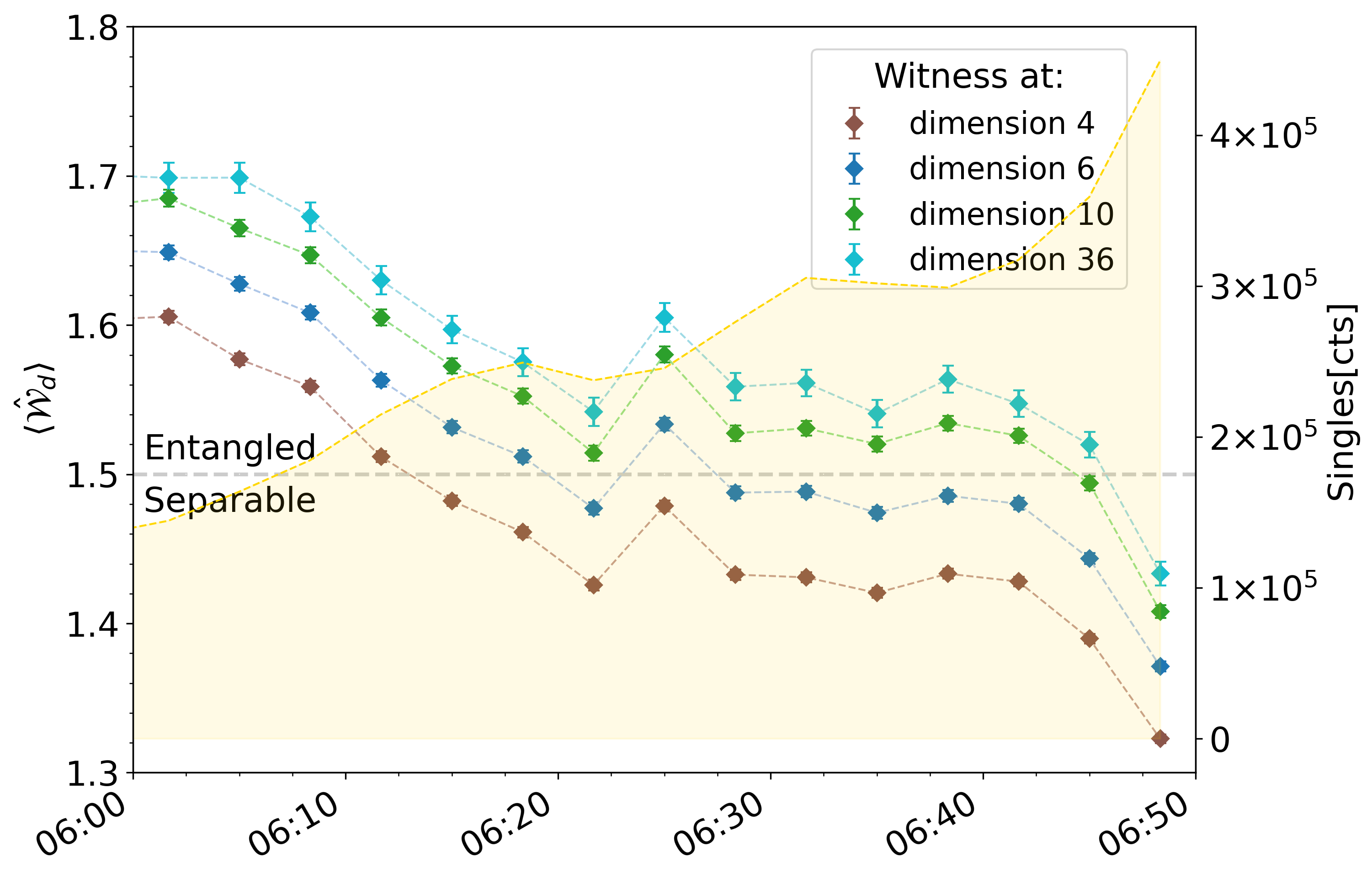}
    \caption{
    \textbf{Entanglement witness.} Expectation values of entanglement witness $\hat{\mathcal{W}}_{d}$ of the discretized distributed state are plotted for different discretization dimensions $d = \{4,6,18,36\}$.
    Values above $1.5$ certify entanglement \cite{spengler}.
    The witness is evaluated for data obtained by integrating over $200\,$s blocks of time-tagged detector event streams, which corresponds to typical passage times of satellites \cite{neumann2018q}. 
    To show the increasing noise from sunlight the received single photon count rates at Bob's side are plotted in the background.
    From the plot it is apparent that with sunrise the channel conditions start to deteriorate rapidly.
    Nevertheless, entanglement can be certified well up to $1.5$ hours after the sunrise (05:03). 
    Further, discretizations with higher  dimensionality  outperform the lower-dimensional ones in all cases.}
    \label{fig:entanglement_witness}
\end{figure}

\subsection{High-dimensional state encoding from discretization}

    Due to the finite precision of recording each individual photon's time of arrival, we implemented discretized measurements in the energy-time DOF.
    These discretized measurements work with notions of \emph{time-frames} and \emph{time-bins}  \cite{ecker2019overcoming,doda2021quantum}.
    A time-frame defines a time interval in which the arrival of up to one photon is expected and its length is one of the free parameters we can choose to manipulate the effective state dimension.
    The time-frame is further subdivided into time-bins which determine whether Alice’s and Bob’s recorded photon arrival times are correlated. 
    One can then specify a time-frame of length $\Delta t_{\mathrm{f}}$ that starts at time $t_0$ and consists of $d$ time-bins of length $\Delta t_{\mathrm{b}}$.
    The measurement outcome in the TOA basis, labeled $j\in \{0,\dots,d-1\}$, indicates the time of arrival in a time-bin starting at time $t_0 + j\Delta t_{\mathrm{b}}$. 
    We use the time of arrival as the standard basis and identify each time-bin to a basis vector. 
    Therefore an outcome $j$ corresponds to the projection onto a state $\ket{j}$.
    The TSUP basis measurement, on the other hand, performs a projection onto the superposition $\frac{1}{\sqrt{2}}\left(\ket{j}\pm\ket{j+k}\right)$, which comes from the indistinguishability between the propagation of entangled photons through the short and the long arms of the MZI.
   Notably, we require that $k = \tau_{\mathrm{MZI}}/\Delta t_{\mathrm{b}}$ to be an integer.
   This ensures that starting times of both $\ket{j}$ and $\ket{j+k}$ are exactly $\tau_{\mathrm{MZI}}$ apart and the interference can be observed.
    The above-mentioned discretization effectively reduces the continuous quantum state introduced in Eq. \eqref{eq:state} to its discretized form
    \begin{align}
          \ket{\Psi}_{AB}\!=\!\sum_{j=0}^{d-1} \frac{1}{\sqrt{d}} \ket{j,j}\!\otimes 
         \!\frac{1}{\sqrt{2}}\left(\ket{H,H}\!+\!e^{-i\phi}\!\ket{V,V}\right).\label{eq:state2}
    \end{align}
    Overall, our measurement setup allows us to measure elements of discretized states of the form in Eq. \eqref{eq:state2}, giving access to $d\times d$-dimensional entangled states.
    In particular, we can choose to address different discretizations by choosing different values of $\Delta t_{\mathrm{f}}$ and $\Delta t_{\mathrm{b}}$, which represent different levels of fine graining.
    This fine-graining process bears the great advantage that it can be adapted and optimized to the various channel and measurement conditions in post-processing.
    In the following, we demonstrate this outstanding feature in an experiment and show that it can be used for optimizing entanglement transfer and secure key rates for a wide range of channel noise contributions under real-world conditions.
    In particular, we are working with a time-frame length of twice the MZI imbalance ($5.4\,$ns), which we divide into multiple time-bins to operate on Hilbert-space sizes varying between $4{\times} 4$ and $36{\times} 36$. 
    Where the choice of $5.4$ ns frame length leads to well defined TSUP measurements (see Method \ref{sub:Methoddiscretization}).
    Furthermore we do not certify entanglement for $d<4$ over the free-space link due to the intrinsic low signal-to-noise ratio \cite{ecker2019overcoming} and $d=36$, corresponding to $\Delta t_b = 150\,$ps, is the highest dimension  achievable beyond which the signal is degraded due to limitations in the timing synchronization and detector jitter(see Appendix \ref{sec:Methods}).

\begin{figure*}
    \centering
    \includegraphics[width=2\columnwidth]{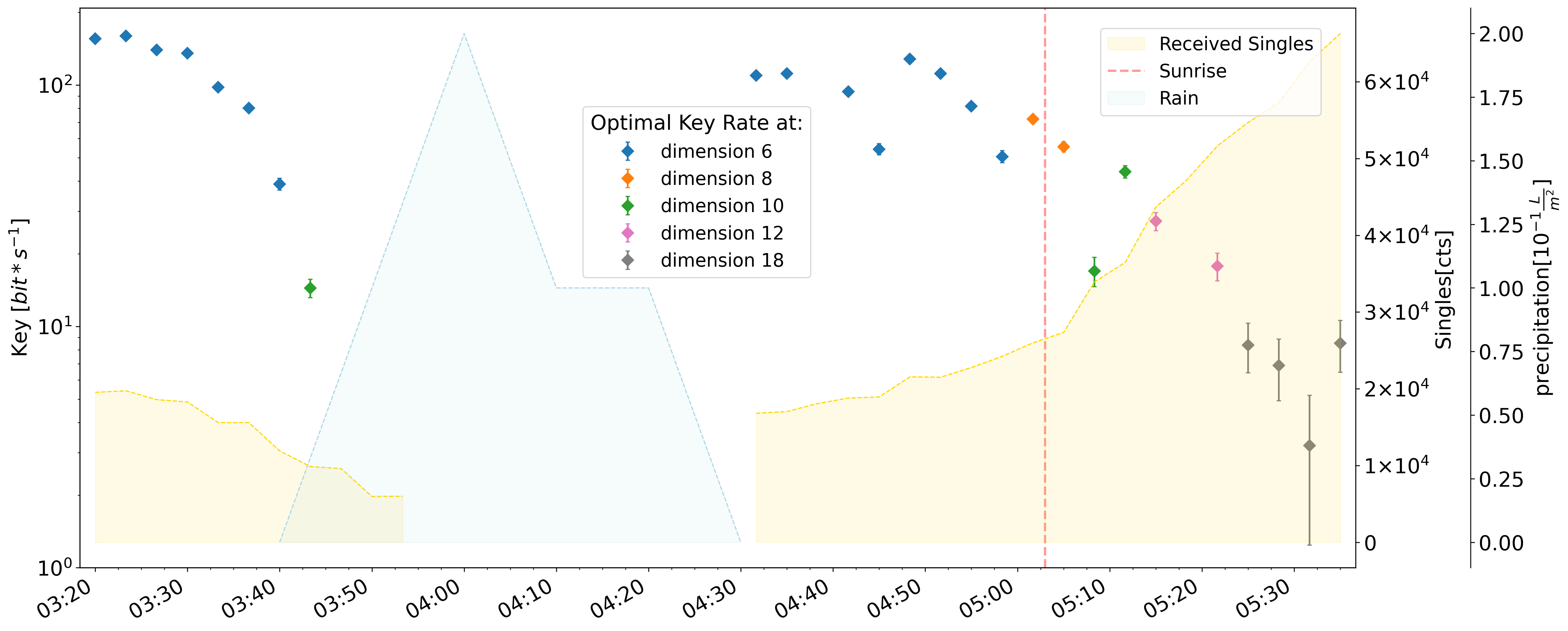}
    \caption{\textbf{Achievable key rate.}The achievable key rates are plotted over the measurement time of the experiment.
    The key rate is evaluated for data obtained by integrating over $200\,$s blocks of time-tagged detector event streams. 
    For each interval the optimal non-zero key-rate value is plotted, using different colors to indicate the corresponding best dimensionality.
    The orange line indicates sunrise and we show the rain data and single photon counts as a measure for background radiation.
    From the plot it can be clearly seen that our noisy free-space channel can be used to distribute a cryptographic key under various noise conditions, even for considerable time after the break of the day.
    Overall, the dimension of the discretization which achieves the largest key rate increases with the increasing background radiation, thus demonstrating superior noise resistance of high-dimensional encoding.
    }
    \label{fig:key_rate}
\end{figure*}
\subsection{High-dimensional entanglement distribution}
    We first demonstrate that we indeed distribute entanglement in the time-energy DOF of the noisy free-space channel.
    For this purpose, we exploit a tight and straightforward entanglement witness $\hat{\mathcal{W}}_d$ that acts on qubit subspaces of the distributed $d$-dimensional state.
    Entanglement is certified when the expectation values $\langle \hat{\mathcal{W}}_d\rangle$ exceed the value of $1.5$ (see \cite{spengler} and Appendix \ref{sub:QubitEntanglementWitness}).
    
    We experienced different types of noise during the $4\,$h measurement run with distinct influences on entanglement distribution and the corresponding optimal state dimensionality.
    The strongest source of noise in the long term is the continuously increasing background radiation from the sun which finally exceeds the level of noise tolerable to observe an entanglement witness at 06:46 (see Fig. \ref{fig:entanglement_witness}).
    Notably, this is more than $1.5$ hours after sunrise  with a noise-count rate on Bob's side that is $20$ times  larger than the signal from the source.
    Other noise types are minor instabilities in the interferometers, caused by vibrations or strong airflow, leading to dephasing noise in the TSUP basis and the dark count rate of the detectors both constantly adding to the noise count.
    Hence, our results show that time-energy entanglement can be successfully distributed in the presence of these different noise contributions over a real-world quantum channel.

\subsection{Noise-resistant time-energy quantum key distribution}

    Here we demonstrate that it is possible  to harness the distributed entanglement for secure quantum communication.
    For this purpose, we employ a recently introduced high-dimensional subspace QKD protocol \cite{doda2021quantum} for which we estimate the achievable asymptotic key rates and demonstrate the advantage of high-dimensional states for quantum communication over urban noisy free-space channels.

   The calculated asymptotic key rates, optimized with respect to dimension $d$, are shown in Fig.\,\ref{fig:key_rate}.
    We clearly observe that the optimal dimensionality changes with the noise conditions in the channel over time. 
    For low noise levels, lower dimensions generate the highest key rates, while higher dimensions show the possibility to obtain a secure key even under the influence of strong noise, particularly during daylight.
    This enables us to adaptively choose the best state dimension, featuring a flexible and versatile approach for QKD under varying noise conditions.
    The optimal dimension strongly depends on the type of noise level the signal is experiencing.
    During precipitation, we observe that the optimal dimension increases (see Fig.\,\ref{fig:key_rate} around 03:45).
    Also during dawn, the ever rising level of background radiation requires an increase of the dimension $d$ to acquire the optimal key (see Fig.\,\ref{fig:key_rate} between 04:30 and 05:35).

\section{\label{sec:Conclusions}Discussion}

    We demonstrated the distribution of high-dimensional entangled photon pairs over a $10.2\,$km-long free-space link under various detrimental noise conditions.
    This was achieved by advancing both experimental techniques and theoretical tools, tailored for the optimal exploitation of the high-dimensional quantum state.
    These results are the first proof of principle demonstration of energy-time entanglement-based QKD over a long-distance free-space channel. The link runs partly over the urban inner-city area which is heavily populated and causes very strong turbulences.
    The developed quantum-state discretization and encoding toolbox (see Appendix \ref{sub:Methoddiscretization}) enables the optimization of QKD under various noise conditions, providing an adaptive and versatile avenue towards noise resistant quantum communication for the future quantum internet.

    The heart of our contribution consists of an innovative way in which we measure and exploit high-dimensional energy-time entanglement.
    This required the first realization of a Franson-type interferometer used over a long-distance free-space link. 
    It was implemented by the deployment of two interferometers and one photon-pair source, all being phase-locked in reference to each other separated by a $10.2\,$km free-space link. Furthermore, we implemented two $4$-f lens systems in the receiver's interferometer to counteract the influence of wave-front distortion that otherwise would inhibit the interferometric measurement.
    For enabling a common reference frame in the time-energy DOF between the communication parties, we had to implement a precise two-step clock synchronization approach which reached a precision of $\sim 20\,$ps.
    These advances break ground for further work in quantum science over long distances and under harsh conditions ranging from fundamental tests, such as the study of Bell non-locality, to advanced applications in quantum information.
    
    Even though the implemented non-local interference measurements only allow for measurements in two-dimensional subspaces, our results provide optimal exploitation of the high-dimensional state space and its applications and benefits. 
    We developed an entanglement-based QKD scheme that effectively utilizes the high-dimensional state space. It can be optimized and adapted concerning the communication channel imperfections, such as strong losses even during rainfall and high solar background radiation. (see Fig. \ref{fig:key_rate}).
    This can be achieved by calculating the state-dimension trade-off between noise resistance and information capacity (see Appendix \ref{sec:Methods}).
    Furthermore, all adaptations can be applied in the post-processing without any change in the setup or knowledge of the noise contribution prior to the measurement.
    Importantly, we also demonstrated that it is necessary to use a state dimension of at least $4$ in our experiment to obtain a positive key rate and that the optimal dimension is often found for even larger total dimensionality, i.e., genuinely requires the usage of more than simple qubit discretizations.%\ru{hehe}{}
    
    We have shown that both entanglement and secure key rates can withstand different noise contributions within the highly turbulent atmospheric channel long after sunrise (see Figs. \ref{fig:entanglement_witness} and \ref{fig:key_rate}). 
    The implemented channel is comparable with LEO satellite links regarding losses and turbulence \cite{marquardt2017quantum,liao2017long}, thus demonstrating its relevance for global quantum communication. 
    Although challenging, it is possible to implement the source and detection modules with the required stabilization and synchronization on space-borne platforms\cite{yin2017satellite}.

    Entanglement even persists further into the day and is still certifiable without any further assumptions, when no key can be produced any more.
    On the one hand this could prove useful for other quantum communication scenarios in which entanglement is always known to provide an advantage, such as entanglement-assisted classical communication \cite{BatmanWinner}.
    On the other hand, this should motivate further research into more elaborate quantum key distribution protocols.
    Since distillable entanglement\cite{ecker2021experimental} can still be shown, one would assume that it should be possible to distill a useful key from the noisy quantum states, despite the fact that current protocols sacrifice too much key to perform error correction to still provide a positive key rate.

    Altogether, our results pave the way for an extensive field of fundamental and applied research in quantum science and technology that explores and exploits the potential of high-dimensional quantum states %\cite{ecker2021remotely}

\section{Acknowledgements}
We thank Matthias Fink, Mirdit Doda and Fabian Steinlechner for their helpful conversations and comments about the experiment. We also thank Robert Blach for the support in setting up the experiment. We also thank the Zentralanstalt für Meteorologie und Geodynamik (ZAMG), especially Roland Potzmann, for making exact weather data available. We also thank the ORF for providing room for the receiver at ORS-Bisamberg.
We acknowledge European Union’s Horizon 2020 programme grant  agreement  No.857156  (OpenQKD)  and  the Austrian  Academy of  Sciences  in  cooperation  with  the FhG  ICON-Programm  “Integrated  Photonic  Solutionsfor Quantum Technologies (InteQuant)”. 
We also gratefully  acknowledge  financial  support  from  the  Austrian Research  Promotion  Agency  (FFG)  Agentur  f\"ur  Luft-und Raumfahrt (FFG-ALR contract 844360 and 854022).
M.H. acknowledge funding from the Austrian Science Fund (FWF) through the START project Y879-N27.
M.P. acknowledges funding and support from VEGA Project~No.~2/0136/19 and GAMU project MUNI/G/1596/2019.

%\section{Author contributions}
%L.B. conceived the experiment under the supervision of R.U.; J.L., S.E. and L.B. developed and implemented the entangled photon source and local detection module; O.K and L.B developed and implemented the laser stabilization. K.H. and L.B. developed and set up the free-space link tracking and receiver optics; J.L. and L.B. developed and deployed the server-client coincidence tracking software; The experiment was carried out by K.H., O.K. , J.L., R.K., S.N., S.E., J.B., M.B. and L.B. under the guidance of R.U.; L.B. performed the data analyses under the guidance of M.P., M.B and M.H.; M.P. and M.H. implemented the entanglement certification methods and the QKD protocol; M.B., M.P. , M.H. and L.B. wrote the first draft; all authors discussed the results and reviewed the manuscript; M.H. and R.U. supervised the whole project. 

\bibliographystyle{apsrev4-1fixed_with_article_titles_full_names_new}
%unsrt}
\bibliography{libary}
\section{\label{sec:Methods}Appendix}
\subsection{Hyper-entangled photon-pair source}
\label{sub:Methodsource}
    The source was built in a Sagnac configuration deploying a $30\,$mm-long type-II ppKTP crystal which was bidirectionally pumped without any active phase stabilization.
    By superposing the clock- and counterclockwise SPDC photons, a polarization-entangled two-photon state was created.
    Further, due to the energy conservation of the SPDC process, energy-time entanglement emerges within the laser’s coherence length,  which leads to the hyperentangled state in Eq.\,\eqref{eq:state}.
    A laser stabilized to a Potassium ($^{39}K$) hyperfine transition pumped the source at $\lambda= 404.453\,$nm with a power of $P=28.5\,$mW and a long-term stability of $\leq $1.8~fm/h.
    The signal and idler photons were produced in a degenerate spectral mode at $\lambda = 808.9\,$nm.
    To evaluate the source performance without link, we carried out local measurements yielding a pair rate of $R_{cc}\sim 65\,$cps/mW with a local heralding efficiency of $\eta\sim 26\, \%$, including detector efficiency and source imperfections.
    Additionally, we measured a local visibility of the entangled state in the horizontal/vertical (diagonal/antidiagonal) polarization basis of $V_{HV}\sim99\,\%$ ($V_{DA}\sim98\,\%$).

\subsection{Detection modules}
\label{sub:MethodDetectionModules}

    The detection modules consisted of two sub-modules, namely the superposition (TSUP) and the time of arrival (TOA) modules, corresponding to the respective measurement basis.
    Both modules received the incoming beam via a beamsplitter to ensure random basis choice.
    Before detection, the TOA basis was further multiplexed in the polarization DOF by means of a polarizing beamsplitter.
    For the TSUP measurement we deployed unbalanced MZIs consisting of two polarizing beamsplitters and a path difference between long and short arm of $\Delta L\sim80.8\,$cm, which corresponds to a propagation-time difference of $\tau_{\mathrm{
    MZI}}\sim 2.7\,$ns.
    The polarizing beamsplitters were used to map the polarization-entangled state to the time-energy entangled state.
    This means that polarization states are mapped to temporal modes as $\ket{V}_{\mathrm{pol}} \rightarrow \ket{L}_{\mathrm{path}}$ and $\ket{H}_{\mathrm{pol}} \rightarrow \ket{S}_{\mathrm{path}}$, where $\ket{S}_{\mathrm{path}}$ and $\ket{L}_{\mathrm{path}}$ represent the short and long paths of the interferometers respectively.
    The measurement after the interferometers was performed in the D/A-basis to delete the photons which-path information.
    This method facilitates a post-selection free Franson interferometer which increases the efficiency of the energy-time analyzers by a factor of $2$.
    Each MZI was locked to a stabilized laser via a piezo actuator in a closed control loop.
    For this purpose, we applied Doppler-free saturation-absorption spectroscopy to lock the lasers \cite{preston1996doppler}.
    For Bob, we used a $780.23\,$nm laser with a $^{87}Rb$ cell ; For Alice, the SPDC pump laser at $404.53\,$nm with a $^{39}K$ cell.
    The stabilization compensated for thermal drifts to keep the phase of the Franson interferometer stable within $1\,$MHz of relative phase error.
    Due to atmospheric wavefront distortion of the incoming beam, Bob's receiving interferometer had two 4f-systems implemented in its long arm. 
    These 4f-systems mapping each incident angle of the arriving beam from the first to the second PBS.
    The 4f-system has been used in the past as spatial filter and for spatial map \cite{vallone2016interference,jin2018demonstration, jin2019genuine}.
    This spatial mapping makes Franson interference possible despite the distorted wavefront.
    The single-photon avalanche diodes (SPADs) are free-space detectors with a detection area of $150\,$\textmu m in diameter on Bob's side and multimode-fiber-coupled SPADs on Alice's side.
    Locally, each Mach-Zehnder interferometer reached an interference visibility of $95\,\%$.

\subsection{Link establishment}
\label{sub:MethodLink}

    The achromatic sending lens ($f=257\,$mm) was mounted on top of a Newtonian telescope ($f=1140\,$mm) with an aperture of $350\,$mm. 
    In addition, a beacon laser ($\lambda=532\,$nm) was mounted on the telescope. 
    On the receiving side, we used a Cassegrain telescope ($f=2032\,$mm) with an aperture of $254$mm in addition to a beacon laser.
    To optimize the pointing, the sending telescope and the entire receiver setup were mounted on Hexapods. 
    This link configuration enabled bidirectional closed-control-loop tracking on the beacon lasers with a $5$\,\textmu rad resolution.
    This method prevents beam wandering and keeps the signal intensity at the optimum.
    To reduce stray light on the receiver, multiple irises were placed in the setup and attached behind the telescope. A combination of longpass, shortpass, and interference bandpass filters was placed in the incoming beam to minimize environmental noise further. The filters were chosen to lower noise in the overall sensitivity spectrum of Si-based detectors by 60 dB.
    With the exception of the band-pass filter with a bandwidth(FWHM) $<1\,$nm at $\lambda= 808.9\,$nm to ensure the narrow spectral filtering of the signal wavelength.

\subsection{Discretization techniques}
\label{sub:Methoddiscretization}

As briefly described in the results section, the crucial component of the discretization is the division of time-tagged detection event streams into time-bins of length $\Delta t_b$.
Subsequently, $d$ subsequent time-bins are collected into a time-frame.
As the quantum state of the photon pairs corresponds to a coherent superposition of arrival times, the information of the arrival time of a photon in one of the $d$ time-bins of a single time-frame forms the computational basis of a $d$-dimensional Hilbert space.
Counting from zero, the arrival in the $j^\mathrm{th}$ time-bin of the time-frame is represented by a basis vector $\ket{j}$.
Using the continuous entangled-state description presented in Eq. \eqref{eq:state}, the energy-time part of the (unnormalized) discretized state can be written  as 
\begin{align}\label{eq:discState}
\ket{\Psi}_{AB} = \sum_{j=0}^{d-1} c_j \ket{j}_A \ket{j}_B,
\end{align}
where $|c_j|= \int_{t_j}^{t_j+\Delta t_b} dt\, f(t)$, $f(t)$ is a continuous function of time, determined by the coherence time of the laser, and $t_j$ denotes the starting time of the $j^\mathrm{th}$ time-bin.
Note that if the coherence time of the laser is much longer than the length of a time-frame $\Delta t_f = |t_{d-1}-t_0+\Delta t_b|$, then $f(t)\approx \bar f=const.$ in this interval which leads to $c_j=\bar f\Delta t_b$ and, after normalization, to $|c_j|=1/\sqrt{d}\quad\forall j$.

In light of the discrete state description above, the TOA measurement can be trivially described as a projection onto the basis $\{\ket{j}\}_{j=0}^{d-1}$.
The basic data unit associated with the corresponding joint measurement is a $d\times d$ correlation matrix $M_{\mathrm{TOA}}=\left(m_{i,j}\right)_{(i,j)}$, which in position $m_{i,j}$ contains the number of coincidences obtained in time-bins $\ket{i}$ and $\ket{j}$ on Alice's and Bob's side, respectively, meaning $m_{i,j} \propto \mathrm{Tr}\left(\hat \rho_{AB}\ket{i}\bra{i}\otimes\ket{j}\bra{j}\right)$.

The description of the TSUP basis measurement is more involved, because only the superposition of arrivals between certain pairs of time-bins can be measured. 
Recall that the delay introduced in the interferometer's long arm is $\tau_\mathrm{MZI}\sim 2.7\,$ns and that the measurement arrangement effectively erases the information about which path (long or short) the photon took.
Therefore, if there are two time-bins $\ket{i}$ and $\ket{j}$ with starting times $t_i$ and $t_j = t_i + \tau_\mathrm{MZI}$ respectively, a click in a TSUP detection module in time interval $[t_{i},t_{i}+\Delta t_b]$ is interpreted as a projection onto an equal superposition of photon arrival in time-bins $\ket{i}$ and $\ket{j}$. 
Since the which-path information is deleted using the D/A-basis measurement in the polarization degree of freedom, the information about the phase between time-bins $\ket{i}$ and $\ket{j}$ in their superposition is encoded in information about which of the two detectors in the TSUP detection module clicked. 
The click of a detector in the transmitted path is interpreted as a projection onto $\ket{+_{i,j}} = \frac{1}{\sqrt{2}}(\ket{i}+\ket{j})$ and the click of a detector in the reflected path as $\ket{-_{i,j}}=\frac{1}{\sqrt{2}}(\ket{i}-\ket{j})$.
The basic data unit for the TSUP measurement therefore contains four $d\times d$ matrices $M_\mathrm{TSUP}^{++}, M_\mathrm{TSUP}^{+-}, M_\mathrm{TSUP}^{-+}$ and $M_\mathrm{TSUP}^{--}$, which capture time-of-arrival information as well as which detector (transmitted $(+)$ or reflected $(-)$ path) in Alice's and Bob's TSUP measurement modules clicked. 
As an example, consider matrix element $m^{+-}_{i,j}$ of $M_\mathrm{TSUP}^{+-}$, which contains the number of coincident projections onto $\ket{+_{i,i'}}$ on Alice's side and $\ket{-_{j,j'}}$ on Bob's side, where $t_{i'} = t_i + \tau_\mathrm{MZI}$ and $t_{j'} = t_j + \tau_\mathrm{MZI}$.
Further complication comes from the fact that not all projections captured in four TSUP correlation matrices belong to the chosen time-frame and therefore their detection is not a valid measurement outcome for the chosen Hilbert space. 
As an example, consider the $d-1^\mathrm{st}$ row of an arbitrary TSUP correlation matrix that contains the number of projections onto a superposition between time-bin $\ket{d-1}$ and a time-bin starting at time $t_{d-1} + \tau_\mathrm{MZI}$ in Alice's lab. Such a projection clearly falls outside of the time-frame defined by time-bins $\ket{0},\dots, \ket{d-1}$ with $t_0 < t_1 < \dots t_{d-1}$.
In this sense the TSUP detection module can project onto an over-complete set of projectors, some of which even fall outside of the chosen discretization.
In order to obtain well-defined and normalized outcome probabilities in the TSUP basis, the set of projectors $\{\ket{+_{i,j}},\ket{-_{i,j}}\}$, which are contained in the four TSUP correlation matrices, need to contain at least one full basis of the Hilbert space associated to the chosen time-frame.

The presented results were obtained by setting the time-frame length to $2\tau_\mathrm{MZI} = 5.4\,$ns. 
Each time-frame is subsequently divided into time-bins of equal size $\Delta t_\mathrm{b}$.
The choice of $\Delta t_\mathrm{b}$ influences the dimensionality $d$ of the Hilbert space as $d=2\tau_{\mathrm{MZI}}/\Delta t_b $. In our experiment, the highest chosen dimensionality was $d = 36$ $(\Delta t_\mathrm{b} = 150\,\mathrm{ps})$ and the smallest one $d = 4$ $(\Delta t_\mathrm{b} = 1350\,\mathrm{ps})$. 
Importantly, if even values of $d$ are used in post-processing, then the four TSUP data matrices contain coincidence counts for projections onto $\{\ket{+_{i,i+d/2}},\ket{-_{i,i+d/2}}\}$ for each $i\in\{0,\dots,d/2-1\}$.
Such a set of vectors forms a complete basis of the chosen $d$ dimensional Hilbert space and thus defines a normalized projective measurement. 
Coincidence counts corresponding to these projectors can be straightforwardly normalized into estimates of outcome probabilities, which are subsequently used in the key rate and entanglement witness calculations.

Having defined the Hilbert spaces and measurements used in the post-processing, it remains to define additional measurement assumptions.
We process the measurement by splitting the data into $200$\,s intervals and sequence them into time-frames. Therefore, both TOA and TSUP matrices used for the calculation of results in different times contain coincident counts from $200$\,s integration of time-frames.
In order to deal with non-conclusive measurement rounds, we employ a fair-sampling assumption. 
In the ideal case, in each of the time-frames exactly one detector would click in both Alice's and Bob's labs -- we call such events \emph{valid time-frames}.
time-frames in which neither party or just a single party detects a photon are discarded, while multi-click time-frames are assigned random outcomes, which are added to the data matrices produced by valid time-frames.

\subsection{Qubit subspace entanglement witness}\label{sub:QubitEntanglementWitness}

As pointed out in \cite{ecker2019overcoming}, strong entanglement robustness can be observed in subspaces of high-dimensional entangled states. 
Therefore, we evaluate a qubit entanglement witness in $2\times 2$ subspaces of the whole $d\times d$ dimensional Hilbert space.
Recall that time-frames of even dimensionality $d$ are used and therefore both Alice's and Bob's local $d$-dimensional Hilbert space can be interpreted as a direct sum of $d/2$ qubit Hilbert spaces $\mathcal{H}_2^{(i)}$, where index $i$ runs from $0$ to $d/2-1$.
Further, each subspace $\mathcal{H}_2^{(i)}$ is spanned by computational basis vectors $\ket{i}$ and $\ket{i+d/2}$. 
In the subspace post-selection, only coincidences where both Alice and Bob registered a click in the same subspace are kept and to each subspace, two measurements are associated. 
The first measurement is a projection onto the time-of-arrival basis spanned by $\{\ket{i},\ket{i+d/2}\}$
and the second one is a projection onto $\{\ket{+_{i,i+d/2}},\ket{-_{i,i+d/2}}\}$.
Both of these form a complete set of projectors in their respective qubit subspaces and thus can be re-normalized into joint outcome probabilities.
Importantly, the employed entanglement witness takes into account only probabilities of matching outcomes in both measurements. 
In each subspace, the  relative frequency of matching outcomes for the TOA measurement is calculated as $p_{\mathrm{TOA}}^i = (m_{i,i}+m_{i+d/2,i+d/2})/(m_{i,i}+m_{i,i+d/2}+m_{i+d/2,i}+m_{i+d/2,i+d/2})$, where $m_{a,b}$ are elements of the raw count matrix $M_{\mathrm{TOA}}$ defined in the previous subsection.
Similarly, the the relative frequency of matching outcomes for the TSUP measurement is calculated as $p_{\mathrm{TSUP}}^i = (m_{i,i}^{++}+m_{i,i}^{--})/(m_{i,i}^{++}+m_{i,i}^{+-}+m_{i,i}^{-+}+m_{i,i}^{--})$, where all $m_{i,i}^{\pm,\pm}$ are elements of the corresponding matrix $M_\mathrm{TSUP}^{\pm\pm}$.
The measured state is entangled \cite{huber2013structure} in subspace $\mathcal{H}_2^{(i)}\times\mathcal{H}_2^{(i)}$ if and only if 
$p_{\mathrm{TOA}}^i+p_{\mathrm{TSUP}}^i > 3/2.$
In Fig. \ref{fig:entanglement_witness}, the average value of the witness calculated over all $d/2$ qubit subspaces is plotted. 
Succinctly, for each subspace $i$ this quantity can be expressed using a witness operator 

\begin{align}
    \hat{\mathcal{W}}_d^{(i)} =& \ketbra{i}{i}\oplus\ketbra{i}{i} +\\\nonumber &\ketbra{i+d/2}{i+d/2}\oplus\ketbra{i+d/2}{i+d/2} +\\\nonumber &\ketbra{+_{i,i+d/2}}{+_{i,i+d/2}}\oplus\ketbra{+_{i,i+d/2}}{+_{i,i+d/2}} +\\\nonumber
    &\ketbra{-_{i,i+d/2}}{-_{i,i+d/2}}\oplus\ketbra{-_{i,i+d/2}}{-_{i,i+d/2}}
\end{align}

with a condition that $\Tr\left(\hat{\rho}^{(i)}_{AB}\hat{\mathcal{W}}_d^{(i)}\right) > 3/2$, where $\hat{\rho}^{(i)}_{AB}$ is the projection of the experimental state $\hat{\rho}_{AB}$ into $i^\mathrm{th}$ subspace.

\subsection{Key-rate estimation}
\begin{figure}[h]
    \centering
    \includegraphics[width=1\columnwidth]{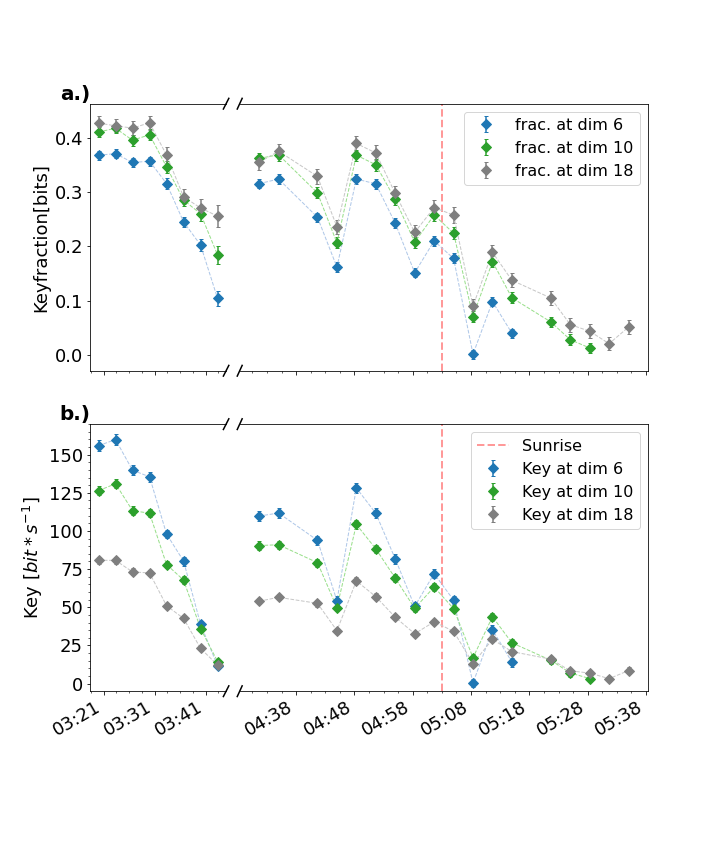}
    \caption{
    \textbf{Key rate vs. Key fraction.} A closer look at achievable asymptotic key rate. In figure \textbf{a.)}, key fraction -- achievable key rate per post-selected coincidence -- is plotted for different values of discretization dimension. 
    To obtain key rate in bits per second, shown in figure \textbf{b.)}, this value is multiplied by the number of subspace coincidences in the given $200$ s integration window. 
    The figure \textbf{a.)} shows that discretizations consistently achieve the highest key fraction with the highest local dimension $d$. 
    However, the highest key fraction does not lead to the overall highest key rate per second in most cases, as using higher-dimensional discretization also leads to a smaller number of subspace coincidences per second. The lines serve as guidance for the eye.
   }
    \label{fig:keyfraction_vs_keyrate}
\end{figure}

For each subspace $\mathcal{H}_2^{(i)}\otimes \mathcal{H}_2^{(i)}$ defined in the previous subsection the asymptotic key-fraction (i.e., key per subspace coincidence) can be estimated using the Koashi-Preskill {\cite{PhysRevLett.90.057902}}{} formula:
\begin{equation}
    K(i) \geq 1 - H(p_{\mathrm{TOA}}^i) - H(p_\mathrm{TSUP}^i),
\end{equation}
\noindent where $H(\cdot)$ is the Shannon entropy function. 
To obtain the overall average key-fraction $K$, individual key-rate fractions $K(i)$ of all subspaces are averaged.
The key rate per second can in turn be calculated by multiplying $K$ by the number of subspace coincidences per second.
Increasing the dimension increases the information shared by each transmitted and accepted photon pair [see Fig. \ref{fig:keyfraction_vs_keyrate} a.)], but on the other hand lessens the number of photon-pair detection events accepted by the protocol and therefore reduces the key rate (see Fig. \ref{fig:keyfraction_vs_keyrate} b.)).
Note that in each integration interval of $200\,$s for which we estimate the key rate (see Fig. \ref{fig:key_rate}), we obtained  coincidence counts ranging from 101k (during night) to 28k (after dawn).

\end{document}